\documentclass[12pt]{article}
\usepackage{amssymb}
\usepackage{amsmath}
\usepackage{amsfonts}
\usepackage{mitpress}
\usepackage{graphicx}

\DeclareMathOperator*{\erfc}{erfc}

\newdimen\dummy
\dummy=\oddsidemargin
\begin{document}

\begin{center}
Pricing of Asian-type and Basket Options via Upper and Lower Bounds

Alexander Novikov\textit{\footnote{%
School of Mathematical and Physical Sciences, University of Technology,
Sydney, Australia, and Steklov Mathematical Institute, Russian Academy of
Sciences. Present address: PO Box 123, Broadway, University of Technology,
Sydney, NSW 2007, Australia; e-mail: Alex.Novikov@uts.edu.au. The work of A.
Novikov (Sections 1,2,3 and 5) was supported by the Russian Scientific Fund,
grant 14-21-00162.}}

Scott Alexander\textit{\footnote{%
School of Mathematical and Physical Sciences, University of Technology, PO
Box 123, Broadway, Sydney, NSW 2007, Australia. The work of S. Alexander
(Section 4) was supported by the Australian Research Council under Grant
DP150102758;}}

Nino Kordzakhia\textit{\footnote{%
Department of Statistics, Macquarie University, North Ride, NSW 2109,
Australia. }}

Timothy Ling\textit{\footnote{%
School of Mathematical Sciences, University of Technology, PO Box 123,
Broadway, Sydney, NSW 2007, Australia. The work of T. Ling (Section 5) was
supported by the Australian Research Council under Grant DP130103315.}}
\end{center}

\textbf{Abstract. }This paper sets out to provide a general framework for
the pricing of average-type options via lower and upper bounds. This class
of options includes Asian, basket and options on the volume-weighted average
price. We demonstrate that in cases under discussion lower bounds allow for
the dimensionality of the problem to be reduced and that these methods
provide reasonable approximations to the price of the option.


\bigskip

1.Introduction. The problem of pricing and hedging Asian-type and basket
exotic options, for more than two underlying assets, has attracted the
considerable attention of many researchers, primarily due to the increasing
demand from practitioners for the development of fast, accurate and stable
numerical procedures suitable for the calculation of prices of
path-dependent options. We believe that methods based on exact lower and
upper bounds meet these requirements. One of the first developments in this
direction for arithmetic Asian call options under geometric Brownian motion
(gBm) was obtained in 1994 by Curran \cite{Cur}.~The method and
corresponding numerical procedure remains widely popular among industry
practitioners to this day. Since 1994 there have been many further advances,
successively leading to more accurate bounds under more complex models for
the underlying asset, such as geometric Levy processes (gLp), stochastic
volatility (SV) and others (see e.g. recent papers \cite{NLK}, \cite%
{BuryakGuo} and references therein). Typically, upper and lower bounds
obtained in these papers have sufficient accuracy to allow them to be used
as approximations for option prices and their sensitivities to parameters.
Contrary to traditional approaches, which are usually based on
finite-difference schemes for solving related partial differential equations
(PDEs), the upper and lower bounds can also be calculated for
high-dimensional problems, within reasonable computational time, while PDE
methods may be unsuitable.

In this paper we provide some extensions to the results on lower and upper
bounds obtained in \cite{NK2014} and demonstrate their applicability to
problems recently discussed in the literature. In Section 3 we discuss the
problem of pricing options on volume-weighted average prices (VWAPs, see
related papers \cite{Stace}, \cite{NLK}, \cite{BuryakGuo}) and develop a new
model for the volume process which is a natural generalisation of the
"Gamma-volume model" suggested in \cite{BuryakGuo}.

In Section 4 we deal with the prices and sensitivities of Asian options
under a geometric Levy process framework. This problem has been approached
by many authors (see e.g. \cite{albrecher_asian_2004}, \cite%
{forde_robust_2010}, \cite{Physic}, \cite{alexander_bounds_2015}). Here for
some special classes of Levy processes we present some new theoretical
results and their numerical implementation based on the use of Fourier
methods.

In Section 5 we suggest a new lower bound for prices of multidimensional
baskets and provide comparisons with recent results obtained in \cite{LaoLo}.

\textbf{2. Notation. A unified approach to deriving lower and upper bounds.}

We assume that the prices of underlying assets $%
S_{t}=(S_{t}^{(1)},....~S_{t}^{(M)}),t\leq T,$ are defined on the filtered
probability space $(\Omega ,\mathcal{F},\{\mathcal{F}_{t}\}_{t\geq 0},%
\mathbb{P})$, with $T$ the maturity time of European-type options. We also
assume that the process $e^{-R_{t}}S_{t}$ is a martingale with respect to
the (risk-neutral) probability measure $\mathbb{P}$ and that $R_{t}~$is the
accrued interest at time $t$. These are standard assumptions in mathematical
finance which are implied by the no-arbitrage hypothesis under a no-dividend
setting (see e.g. \cite{Shir} concerning other details of this approach).
This may be easily extended to the case where dividends are paid.

We aim to obtain accurate bounds for prices of Asian-type and basket options
which, in a most general form, can be written as

\begin{equation}
C_{T}=\mathbb{E}e^{-R_{T}}\Big(\int_{0}^{T}\big(Y_{u}-K\big)\mu (du)\Big)%
^{+},  \label{eqn:AsianBasketPrice}
\end{equation}%
where $x^{+}=\max [x,0],~K~$is a constant ("strike"), $Y=(Y_{u},0\leq u\leq
T)$ is a random process adapted to the filtration and $\mu (u)$ is a
(generally speaking) random distribution on $[0,T]$. As examples, one may
consider the following financial options which occur often in practice. By
taking $Y_{u}=\sum_{i=1}^{N}w_{i}S_{u}^{(i)},~$with $w=(w_{i},i=1,...,N)~$a
given vector of weights and concentrating $\mu (u)$ at the point $u=T,~$ we
obtain  the price of a standard basket-call option

\begin{equation}
C_{T}=\mathbb{E}e^{-R_{T}}\bigg(\sum_{i=1}^{N}w_{i}S_{T}^{(i)}-K\bigg)^{+};
\label{BASKET}
\end{equation}%
by setting $\mu (du)=du/T,$ we obtain the price of a continuously-monitored
(CMO) Asian basket call; by taking $\mu (du)=dV_{u}/V_{T}$~with $V_{u}~$ the
accumulated total traded volume at the moment $u$, we obtain the case of
options on the volume-weighted average price (VWAP). The notation can also
accommodate the case of discretely monitored options (DMO) on VWAP:

\begin{equation}
C_{T}=\mathbb{E}e^{-R_{T}}\bigg(\frac{\sum\limits_{t_{j}\leq
T}S_{t_{j}}U_{t_{j}}}{\sum\limits_{t_{j}\leq T}U_{t_{j}}}-K\bigg)^{+},
\label{DMO VWAP}
\end{equation}%
where $U_{t_{j}}$ is the traded volume at the moment $t_{j}$.

We also use the following convenient notation

\begin{equation}
\left\langle H,\mu \right\rangle :=\int_{0}^{T}H_{u}\mu (du),~H\in \mathcal{H%
}^{\mu },  \label{h}
\end{equation}%
where we suppose that the integral is well defined (e.g.$~\mathcal{H}^{\mu
}:=L^{1}(\Omega \times \lbrack 0,T],\mathcal{F\otimes }~\mathbb{B[}0,T],d%
\mathbb{P}\times d\mu )$\footnote{%
5This notation was suggested by a reviewer}.

Using this notation we can rewrite (\ref{eqn:AsianBasketPrice}) in the more
compact form:

\begin{equation*}
C_{T}=\mathbb{E}e^{-R_{T}}\left\langle Y-K,\mu \right\rangle ^{+}=\mathbb{E}%
e^{-R_{T}}(\left\langle Y,\mu \right\rangle -K)^{+}.
\end{equation*}

The following result is, in fact, a generalisation of Theorem 1 from \cite%
{NK2014} which was proved under slightly different settings and notation.

Let $\mathbb{I}\{A\}~$be the indicator of the set $A$.

\textbf{Proposition 1.} \textit{Let }$z\in \mathbb{R},Y\in \mathcal{H}^{\mu
},H\in \mathcal{H}^{\mu }.$\textit{~Then}%
\begin{align}
C_{T}& =\max_{z\in \mathbb{R},H\in \mathcal{H}^{\mu }}\{\mathbb{E}%
e^{-R_{T}}\left\langle Y-K,\mu \right\rangle \mathbb{I}\{\left\langle H,\mu
\right\rangle >z\}\},  \label{plus0} \\
C_{T}& =\min_{H\in \mathcal{H}^{\mu }}\{\mathbb{E}e^{-R_{T}}\left\langle
(Y-K+\left\langle H,\mu \right\rangle -H)^{+},\mu \right\rangle \},
\label{plus1}
\end{align}%
\textit{where both the maximum and the minimum are attained by taking}$~z=K$%
\textit{, }$H=Y$\textit{.}

These relations can be used for obtaining lower and upper bounds for $C_{T}$
by choosing a suitable random variable $H$ (we call it a "proxy", see
examples below).

There exist many other similar relations, e.g. with arbitrary random
variables $X$ in (\ref{plus0}) instead of $\left\langle H,\mu \right\rangle $
but we would like to stress here that the maximum and the minimum in
Proposition 1 are both attained by setting $H=Y$. To some extent relation (%
\ref{plus1}) resembles one of the famous dual representations for American
options suggested by L. C. G. Rogers \cite{Rogers}.

For completeness of the exposition we provide the proof of \textbf{%
Proposition 1}.

\textbf{Proof.} For any$~z\in \mathbb{R}\mathcal{~}$and $H\in \mathcal{H}%
^{\mu }$%
\begin{equation*}
\left\langle Y-K,\mu \right\rangle ^{+}\geq \left\langle Y-K,\mu
\right\rangle \mathbb{I}\{\left\langle H,\mu \right\rangle >z\}.
\end{equation*}%
Thus we obtain%
\begin{equation}
C_{T}=\mathbb{E}e^{-R_{T}}\left\langle Y-K,\mu \right\rangle ^{+}\geq
\sup_{z\in \mathbb{R},H\in \mathcal{H}^{\mu }}\{\mathbb{E}%
e^{-R_{T}}\left\langle Y-K,\mu \right\rangle \mathbb{I}\{\left\langle H,\mu
\right\rangle >z\}\}.  \notag
\end{equation}%
Since $\left\langle Y-K,\mu \right\rangle ^{+}=\left\langle Y-K,\mu
\right\rangle \mathbb{I}\{\left\langle Y,\mu \right\rangle >K\},~$ the
equality in (\ref{plus0}) is attained when $z=K$ and $H=Y$.

To prove (\ref{plus1}) we note that for any $H\in \mathcal{H}^{\mu }$%
\begin{align}
\left\langle Y-K,\mu \right\rangle ^{+}& =\left\langle Y-K+\left\langle
H,\mu \right\rangle -H,\mu \right\rangle ^{+}  \notag \\
& \leq \left\langle (Y-K+\left\langle H,\mu \right\rangle -H)^{+},\mu
\right\rangle ,  \label{RHS1}
\end{align}%
where the last inequality is due to convexity of the function $x^{+}$. This
implies that 
\begin{equation*}
C_{T}\leq \inf_{H\in \mathcal{H}^{\mu }}\{\mathbb{E}e^{-R_{T}}\left\langle
(Y-K+\left\langle H,\mu \right\rangle -H)^{+},\mu \right\rangle \}.
\end{equation*}%
Note that when $H=Y\ $ we have%
\begin{equation*}
\mathbb{E}\left\langle Y-K+\left\langle Y,\mu \right\rangle -Y,\mu
\right\rangle ^{+}=\mathbb{E}(\left\langle Y,\mu \right\rangle -K)^{+}=C_{T}.
\end{equation*}%
Hence the minimum in (\ref{plus1}) is achieved when $H=Y$ and (4) does
indeed hold. Thus the proof of Proposition 1 has been completed.

For the remainder of the paper we shall consider only the case where $%
R_{t}=rt$ with $r$ a constant. The case of time-dependent deterministic
interest rates can be easily adapted to our framework (see related examples
in \cite{NK2014}).

Due to (\ref{plus0}) from Proposition 1 we have that for any $H\in \mathcal{H%
}^{\mu }$ 
\begin{equation}
C_{T}\geq LB:=\sup_{z\in \mathbb{R}}\{\mathbb{E}e^{-rT}\left\langle Y-K,\mu
\right\rangle I\{\left\langle H,\mu \right\rangle >z\}\}.  \label{LB}
\end{equation}%
The evaluation of the lower bound $LB$ is typically much simpler than that
of the true price $C_{T}$ due to the fact that (\ref{LB}) does not contain
the non-linear function $x^{+}$. As a result, one may note that for the
evaluation of $LB$ we need only the joint distribution of $\ln (Y_{t})$~and~$%
\left\langle H,\mu \right\rangle ,~$see examples in Sections 3, 4 and 5.

By (\ref{plus1}) from Proposition 1 we have for any $H\in \mathcal{H}^{\mu }$
and any $a\in \mathbb{R}$%
\begin{equation*}
C_{T}\leq e^{-rT}\mathbb{E}\left\langle (Y-K+a\left\langle H,\mu
\right\rangle -aH)^{+},\mu \right\rangle 
\end{equation*}%
and hence%
\begin{equation}
C_{T}\leq UB:=e^{-rT}\inf_{a\in \mathbb{R}}\{\mathbb{E}\left\langle
(Y-K+a\left\langle H,\mu \right\rangle -aH)^{+},\mu \right\rangle \}.
\label{UB}
\end{equation}

The evaluation of $UB~$is easier when compared to that of $C_{T}~$but it is
typically more time consuming than the evaluation of the $LB$ (\ref{LB})
because it$~$does involve the non-linear function $x^{+}$.

A discussion concerning reasonable choices of the proxy random process $%
H^{\ast }$ in (\ref{LB}) and (\ref{UB}) as it relates to the production of
accurate bounds is presented in \cite{NK2014}. We would like to remind
readers that the maximum and the minimum in Proposition 1 are both attained
by setting $H=Y$ and, therefore, one could expect that if we have a "good"
approximation $H^{\ast }$ for $Y$ (i.e. leading to accurate $LB$ in (\ref{LB}%
)) then choosing the same proxy $H^{\ast }$ in (\ref{UB}) will produce a
"good" $UB$ in (\ref{UB}). This consideration is often supported by
comparisons with results of Monte Carlo simulations (see e.g. \cite{NK2014}, 
\cite{alexander_bounds_2015})) but, of course, it would be desirable to have
an analytical estimate of the distance between $UB$ and $LB$. In most
examples the accuracy of LB, when compared with Monte Carlo simulation, is
so good that such $LBs$ can even be used to approximate sensitivities
(Greeks) via numerical differentiation (see examples in Section 4).

\textbf{3. Lower bound for options on VWAP under Levy-volume model.}

In this Section we discuss only the one-dimensional case and set $%
Y=S=(S_{t},0\leq t\leq T)$.~We shall assume that the process$~S$ and the
accumulated traded volume $V=(V_{t},0\leq t\leq T)~$are independent.

\textbf{3.1 Levy-volume model. \ }In \cite{NLK} we applied the method of
matching moments and log-normal approximation for $\int_{0}^{T}S_{u}dV_{u}~$%
under the assumptions that $S$ is a geometric Brownian motion (gBm) and that
the accumulated volume process $V$ is the integral of a squared
Ornstein-Uhlenbeck $U=(U_{t},0\leq t\leq T)$ $V_{t}=\int_{0}^{t}U_{u}^{2}du$%
).

The key point of our approach in \cite{NLK} was the development of a
technique for finding the function%
\begin{equation*}
g=\Big(g_{t}:=\mathbb{E}\frac{U_{t}^{2}}{V_{T}},~t\leq T\Big).
\end{equation*}%
The method of finding $g$ proposed in \cite{NLK}, is based on the formula 
\begin{equation}
g_{t}=-\int_{0}^{\infty }\frac{\partial }{\partial z}\mathbb{E}%
e^{-zU_{t}^{2}-qV_{T}}\big|_{z=0}dq,  \label{g}
\end{equation}%
which, as a matter of fact, holds for any random variable $U_{t}$ and any
positive random variable $V_{t}.~$ Relation (\ref{g}) leads to analytical
representations for moments of VWAP and bounds for $C_{T}$ (see details in 
\cite{PhD thesis of TL}, \cite{NK2014}, \cite{NLK} ).

Here we discuss another model, that we call a Levy-volume model, which is a
generalisation of the Gamma-volume model suggested by Buryak and Guo \cite%
{BuryakGuo}. The basic assumption of our model is that the increments of
accumulated trading volume are increments of a Levy subordinator $%
V=(V_{t},t\leq T)$, i.e. a process $V_{t}\ $with independent and stationary
increments,~ $V_{0}=0$, $\mathbb{P}(V_{t}>0)=1$ for $t>0$. Under this model
for any trading moments $t_{j}$ and $t_{j-1}$~we define increments of
trading volumes as follows: 
\begin{equation*}
U_{t_{j}}:=V_{t_{j}}-V_{t_{j-1}},~0=t_{0}<t_{1}<...<t_{N}=T.
\end{equation*}%
Note that in \cite{BuryakGuo} it was assumed that $V_{t}$ is a Gamma-process
with a marginal gamma-density function%
\begin{equation}
\mathbb{P}(V_{t}\in dx)=\beta (\beta x)^{\alpha t-1}\frac{e^{-\beta x}dx}{%
\Gamma (\alpha t)},~\alpha >0,~\beta >0,~t>0,~x>0.  \label{gamma}
\end{equation}%
We stress here that, contrary to the model with a squared Ornstein-Uhlenbeck
process in \cite{NLK} and \cite{PhD thesis of TL}, the accumulated
Levy-volume process $V_{t}$ has independent increments and, apparently, this
leads to a simplification of the calculation of lower bounds for DMO and CMO
on VWAP options.

In the CMO version of the Levy-volume model it is natural to set%
\begin{equation*}
\mu (du)=\frac{dV_{u}}{V_{T}},~0\leq u\leq T,
\end{equation*}%
and in the DMO version we shall assume that $\mu (u)$ has atoms at trading
times $\{t_{j},j=1,...,N\}$ and hence use (\ref{DMO VWAP}) for finding $C_{T}
$.

For further calculations the following simple fact plays an important role.

\textbf{Proposition 2.} Let \textit{\ }$V_{t}$\textit{\ }be a subordinator.
Then%
\begin{equation}
f(u):=\mathbb{E}\frac{V_{u}}{V_{T}}=\frac{u}{T}~,~u\leq T.  \label{f}
\end{equation}

The proof of this fact is elementary when $u/T=i/N$ is a rational number.
Indeed, we can then write $V_{u}=\sum\limits_{j=1}^{i}\xi _{j}$, where
random variables $\xi _{j}:=V_{jT/N}-V_{(j-1)T/N}$ are independent and
identically distributed. Since by distribution $(\xi _{j},V_{T})=(\xi
_{1},V_{T})$ we have

\begin{equation*}
\mathbb{E}\frac{V_{u}}{V_{T}}=\sum\limits_{j=1}^{i}\mathbb{E}\frac{\xi _{j}}{%
V_{T}}=i\mathbb{E}\frac{\xi _{1}}{V_{T}},~i=1,2,\ldots ,N.
\end{equation*}%
This implies when $i=N~$that $N\mathbb{E(}\xi _{1}/V_{T})=1$ and therefore$~i%
\mathbb{E(}\xi _{1}/V_{T})=i/N=u/T.$

The case of arbitrary $u$ and $T$ can be treated \ with a limit procedure or
with the help of formula (\ref{g}) or with the following considerations. It
is easy to see that $f(u):=\mathbb{E}V_{u}/V_{T}$ is a nondecreasing
function, $f(0)=0,$ $f(T)=1$ and for any$~u\in \lbrack 0,T],~s\in \lbrack
0,T]~$ such that $u+s$~$\in \lbrack 0,T]$%
\begin{align*}
f(u+s)& =\mathbb{E}\frac{V_{u}+V_{u+s}-V_{u}}{V_{T}}=f(u)+\mathbb{E}\frac{%
V_{u+s}-V_{u}}{V_{T}} \\
& =f(u)+f(s),~
\end{align*}%
where the last equality is due to stationarity and independence of
increments of $V_{t}$. It is well known that the monotonic solution of the
functional equation $f(u+s)=f(u)+f(s)$ is a linear function, and since $%
f(0)=0,$ $f(T)=1~$we have  $f(u)=u/T$.

\textbf{3.2 Lower bounds.} First we note that due to the convexity of the
function $x^{+}$ and the assumption of independence of $V$ and $S_{t}$ (the
dynamics of $S_{t}$ have not yet been specified), we always have the lower
bound of the form%
\begin{equation*}
C_{T}:=e^{-rT}\mathbb{E}\left\langle S-K,\mu \right\rangle ^{+}\geq e^{-rT}%
\mathbb{E}\left\langle S-K,\mathbb{E}\mu \right\rangle ^{+}
\end{equation*}%
where, due to Proposition 2 under the Levy-volume model, we have for the CMO
case 
\begin{equation*}
\mathbb{E}\mu (du)=du/T.
\end{equation*}
This means that we can obtain $LBs$ for options on VWAP via $LBs$ for
ordinary Asian options.

We have already demonstrated how our technique works for deriving $LBs$ of
both CMO and DMO for Asian options under a Gaussian framework in \cite%
{NK2014}, but for completeness of the exposition here we demonstrate some
steps of this technique.

To allow for the comparison of results we assume further, as in \cite%
{BuryakGuo}, that $S_{t}$ is a gBm process. Recall that under the
risk-neutral measure we have the representation%
\begin{equation*}
S_{t}=S_{0}\exp \{(r-\frac{\sigma ^{2}}{2})t+\sigma B_{t}\},
\end{equation*}%
where $B=(B_{t},t\leq T)$~is a standard Brownian motion and $\sigma $ is the
volatility. (Note that a generalisation to the case of geometric Levy models
can be carried out with a technique reported in Section 4 below). For the
CMO case we suggest to use $\left\langle \ln (S),\mu \right\rangle ~$as the
proxy $\left\langle H,\mu \right\rangle ~$in (\ref{plus0}).\ Under the
assumptions listed above this leads to the following representation for $LB$
in the CMO case:

\begin{equation}
LB^{C}:=e^{-rT}S_{0}\sup_{z}\mathbb{E}\bigg(\int_{0}^{T}e^{(r-\frac{\sigma
^{2}}{2})u+\sigma B_{u}}\frac{du}{T}-K\bigg)\mathbb{I}\bigg{\{}%
\int_{0}^{T}B_{t}\frac{du}{T}>z\bigg{\}}.  \label{LBCMO}
\end{equation}%
Similarly, in the DMO case we use the proxy $\frac{1}{N}%
\sum_{j=1}^{N}B_{t_{j}}$ and thus we have%
\begin{equation*}
LB^{D}:=e^{-rT}S_{0}\sup_{z}\bigg(\sum_{j=1}^{N}\mathbb{E}e^{(r-\frac{\sigma
^{2}}{2})t_{j}+\sigma B_{t_{j}}}\mathbb{I}\bigg{\{}\frac{1}{N}%
\sum_{j=1}^{N}B_{t_{j}}>z\bigg{\}}\frac{t_{j}-t_{j_{-1}}}{T}
\end{equation*}%
\begin{equation}
-K\mathbb{P}\bigg(\frac{1}{N}\sum_{j=1}^{N}B_{t_{j}}>z\bigg)\bigg).
\label{LB1Da}
\end{equation}%
The expressions (\ref{LBCMO}) and (\ref{LB1Da}) have already been discussed
in \cite{NK2014} under a setting that includes general Gaussian processes.
In particular, using the Girsanov transformation in the case of gBm we have
managed to make further simplifications to (\ref{LBCMO}) and obtained the
following lower bound for the CMO case:%
\begin{align*}
LB^{C}& =e^{-rT}\sup_{z}\bigg(\int_{0}^{T}S_{0}e^{ru}\erfc(\frac{%
z-\sigma u(1-u/(2T))}{\sqrt{2T/3}})\frac{du}{T} \\
& \qquad \qquad \qquad -K\erfc\bigg(\frac{z}{\sqrt{2T/3}}\bigg)\bigg),
\end{align*}%
where $\erfc(x):=\frac{2}{\sqrt{\pi }}\int_{x}^{\infty }e^{-t^{2}}dt$ and 
$x\in \mathbb{R}$. In the DMO case with $t_{i}=iT/N$, (\ref{LB1Da}) can be
reduced to 
\begin{align}
LB^{D}=\frac{e^{-rT}}{2T~N}\sup_{z}\bigg(& \sum_{i=1}^{N}e^{rt_{i}}S_{0}\erfc%
\Big(z-\sigma t_{i}\Big(T-\frac{t_{i}}{2}+\frac{T}{2N}\Big)\Big)
\label{LB^D} \\
& -K\erfc\big(z(2D_{N})^{-1/2}\big)\bigg),  \notag
\end{align}%
where $D_{N}=\frac{T}{3}(1+\frac{3}{2N}+\frac{1}{2N^{2}})~$ (see \cite%
{NK2014}).

\textbf{3.3 Numerical results. }The following table presents values of the
lower bounds $LB^{D}~$(\ref{LB^D}) and the log-normal approximations $LogNorm
$ based on the approximation of variance for VWAP from (\cite{BuryakGuo},
equation (5.5)). Note that the value $3.0120^{\ast }$ is from (\cite%
{BuryakGuo}), Table 3). We also provide Monte Carlo estimates ($MC$) based
on 10$^{7}$ trajectories for the case discussed in (\cite{BuryakGuo}). i.e.
for a Levy-volume model generated by a Gamma-process $V_{t}$~with parameters 
$\alpha =10$,~ $\beta =1$ in (\ref{gamma}). Standard mean-square deviations
of $MC~$estimates in Tables 1, 4 and 5 are denoted as $SE.$

As in (\cite{BuryakGuo}) we set $S_{0}=100$, $K=100$, $r=0.05$, $N=80$ and $%
T=0.317.$

\bigskip

\begin{table}[h]
\caption{ Gamma-volume model}
\label{tbl:Novikov}\centering
\begin{tabular}{lcccc}
 $\sigma$ & $LB^D$ & $LogNorm$ & $MC$ & $SE$ error \\ 
 0.2 & 3.0057 & 3.0120* & 3.0056 & 0.0013 \\ 
0.4 & 5.5570 & 6.0170 & 5.5866 & 0.0027 \\ 
0.6 & 8.1130 & 8.5930 & 8.1542 & 0.0042 \\ 
0.8 & 10.6580 & 11.1680 & 10.7190 & 0.0059 \\ 
\end{tabular}%
\end{table}

The results from Table 1 confirm our observation made in \cite{NLK}, \cite%
{NK2014}, \cite{PhD thesis of TL} that the log-normal approximation is not
accurate enough when values of $\sigma ~$(volatilities) are relatively
large. Meanwhile the relative errors for $LB^{D}$ are always less than $%
0.5\% $.

\textbf{4. Asian options under Levy process framework.}

In this Section we discuss only some examples of numerical evaluations of
lower bounds for Asian options on a one-dimensional asset $S$.~We shall
assume that the process$~S$ is modeled as a geometric Levy process.

\textbf{4.1 Bounds for options with fixed strikes. }Let $Z:=(Z_{t}=\ln
(S_{t}/S_{0}),~0\leq t\leq T)$ and the interest rate $r~$be constant.
Assuming in (\ref{plus0}) $Y=S$ and using the proxy $H=Z$ we obtain the
following formula for Asian options 
\begin{equation}
C_{T}\geq LB^{A}:=e^{-rT}\sup_{z\in \mathbb{R}}\mathbb{E}\left\langle
S_{0}e^{Z}-K,\mu \right\rangle \mathbb{I}\{\left\langle Z,\mu \right\rangle
>z\}.  \label{eqn:AlexanderLB}
\end{equation}%
Similarly, using the proxy $H=aZ$, $a\in \mathbb{R}$ in (\ref{plus1}) we
have 
\begin{equation*}
C_{T}\leq UB^{A}:=e^{-rT}\inf_{a\in \mathbb{R}}\mathbb{E}\left\langle
(S_{0}e^{Z}-K+a\left\langle Z,\mu \right\rangle -aZ)^{+},\mu \right\rangle .
\end{equation*}%
This upper bound has been investigated in \cite{NK2014} under the  gBm
framework. In this paper we shall consider several examples under the gLp
framework and find the lower bound $LB^{A}$ which is typically more accurate
and simpler to calculate than $UB^{A}$.

\textbf{4.2 Computing Lower bounds under the Levy process framework with
Fourier transforms. }We assume that $Z=(Z_{t}=\ln (S_{t}/S_{0}),~0\leq t\leq
T)$ is a Levy process. We also assume that the joint characteristic function
of $Z_{t}~$and $\left\langle Z,\mu \right\rangle $ 
\begin{equation*}
\varphi (\xi ,\zeta ;t):=\mathbb{E}e^{i\xi Z_{t}+i\zeta \left\langle Z,\mu
\right\rangle }
\end{equation*}%
is known in an explicit form (see examples below). We assume also that there
exists the density function $f_{\left\langle Z,\mu \right\rangle }(z)$,
calculated as the inverse Fourier transform of $\varphi (0,\zeta ;T)$.

The lower bound $LB^{A}~$can be calculated in two steps. In the first step
we find the value of $z=\underline{z}~$ that gives the global maximum in (%
\ref{eqn:zOptimalFixed}); in the second step we use $\underline{z}$ to the
maximum that is $LB^{A}$.

The equation for $\underline{z}$ can be found by the following way.
Differentiating over $z~$the expression under the symbol $\sup ~$in (\ref%
{eqn:AlexanderLB}) \ and equating to zero we have%
\begin{equation*}
\frac{d}{dz}\int_{0}^{T}\mathbb{E}e^{Z_{u}}\mathbb{I}\{\left\langle Z,\mu
\right\rangle >z\}\mu (du)=\frac{K}{S_{0}}f_{\left\langle Z,\mu
\right\rangle }(\underline{z}).
\end{equation*}%
Then using Fourier techniques on the conditional expectation (in a similar
way to \cite{Physic}), and noting various symmetries in the integrand we
obtain%
\begin{equation}
\frac{1}{\pi }\int_{0}^{T}\int_{0}^{\infty }\Re \big(\varphi (-i,\zeta
;u)e^{-i\zeta \underline{z}}\big)d\zeta \mu (du)=\frac{K}{S_{0}}%
f_{\left\langle Z,\mu \right\rangle }(\underline{z}).
\label{eqn:zOptimalFixed}
\end{equation}%
Having $\underline{z}$ the lower bound $LB^{A}$ can be reduced to the
calculation of the following integral:\textbf{\ }for $\alpha _{1}>0$, $%
\alpha _{2}<-1$ and $\beta <0$ 
\begin{equation}
LB^{A}:=\frac{e^{-R_{T}}S_{0}}{2\pi ^{2}}\int_{0}^{T}\int_{-\infty }^{\infty
}\int_{0}^{\infty }\Re \big(\hat{h}_{1}(\xi ,\zeta ;\underline{z})\varphi
(-\xi +i\alpha _{1},-\zeta +i\beta ;u)  \label{1234}
\end{equation}%
\begin{equation*}
+\hat{h}_{2}(\xi ,\zeta ;\underline{z})\varphi (-\xi +i\alpha _{2},-\zeta
+i\beta ;u)\big)d\xi d\zeta \mu (du),
\end{equation*}%
where 
\begin{align*}
\hat{h}_{1}(\xi ,\zeta ;\underline{z})& =\bigg(\frac{K}{(\alpha _{1}+i\xi
)S_{0}}-\frac{1}{\alpha _{1}+1+i\xi }\bigg)\frac{e^{\underline{z}(\beta
+i\zeta )}}{\beta +i\zeta }, \\
\hat{h}_{2}(\xi ,\zeta ;\underline{z})& =\bigg(\frac{1}{\alpha _{2}+1+i\xi }-%
\frac{K}{(\alpha _{2}+i\xi )S_{0}}\bigg)\frac{e^{\underline{z}(\beta +i\zeta
)}}{\beta +i\zeta }.
\end{align*}

The derivation of (\ref{1234}) can be done in two steps. First using the
relations (\ref{plus0}) and (\ref{eqn:zOptimalFixed}) we obtain 
\begin{align}
LB^{A}& :=e^{-rT}\mathbb{E}\left\langle S_{0}e^{Z}-K,\mu \right\rangle 
\mathbb{I}\{\left\langle Z,\mu \right\rangle >\underline{z}\}  \notag \\
& =e^{-R_{T}}S_{0}\int_{0}^{T}\big(\Psi _{1}(u;\underline{z})+\Psi _{2}(u;%
\underline{z})\big)\mu (du),  \label{eqn:AlexanderLBproof}
\end{align}%
where 
\begin{equation*}
\Psi _{1}(u;\underline{z}):=\int_{-\infty }^{\infty }\int_{-\infty }^{\infty
}\Big(\Big(e^{x}-\frac{K}{S_{0}}\Big)\mathbb{I}\{x<0,y>\underline{z}\}\Big)%
f(x,y;u)dxdy,
\end{equation*}%
\begin{equation*}
\Psi _{2}(u;\underline{z}):=\int_{-\infty }^{\infty }\int_{-\infty }^{\infty
}\Big(\Big(e^{x}-\frac{K}{S_{0}}\Big)\mathbb{I}\{x>0,y>\underline{z}\}\Big)%
f(x,y;u)dxdy,
\end{equation*}%
and$~f(x,y;t)$ is the joint density of $(Z_{t},\left\langle Z,\mu
\right\rangle )$. Then for the second step we use the exponential-damp (see 
\cite{borovkov_new_2002}) which leads to (\ref{1234}).

\textbf{4.3 Numerical results. }Tables 2 and 3 display the lower bounds and
approximations for the sensitivity with respect to $S_{0}~$("Delta")
respectively for a variety of processes (variance-gamma (VG),
normal-inverse-Gaussian (NIG), Merton jump-diffusion (JD), see details in 
\cite{alexander_bounds_2015}). We assume $S_{0}=K=100$, the monitoring
points are equally-spaced, time $T=1$ is measured in years and $N=\infty $
indicates a continuously-monitored option (CMO). The Monte Carlo estimates ($%
MC$) are based on 10$^{7}$ trajectories.

\bigskip

\begin{table}[h]
\caption{ Lower bounds $(LB^{A})$ and $(MC)$ results}
\label{tbl:AlexanderLB}\centering
\begin{tabular}{lcccccc}
$X$ process & Type & $T$ (years) & $N$ & MC & $LB^A$ &  \\ 
VG & Fixed & 1 & 10 & 6.2150 & 6.2118 &  \\ 
NIG & Fixed & 1 & 20 & 5.9855 & 5.9770 &  \\ 
Merton & Fixed & 1 & $\infty$ & 5.7730 & 5.7634 &  \\ 
\end{tabular}%
\end{table}

\begin{table}[h]
\caption{Delta approximations and $MC$ results }
\label{tbl:AlexanderDelta}\centering
\begin{tabular}{lcccccc}
 $X$ process & Type & $T$ (years) & $N$ & MC & Delta &   \\ 
VG & Fixed & 1 & 10 & 0.5943 & 0.5943 &   \\ 
NIG & Fixed & 1 & 20 & 0.5912 & 0.5914 &    \\ 
Merton & Fixed & 1 & $\infty$ & 0.6084 & 0.6082 &    \\ 
\end{tabular}%
\end{table}

Comparison of the LB results to MC estimates shows that the lower bound is
very accurate and for most purposes may be taken to be the actual price.

\bigskip \textbf{5. Lower bound for Baskets options}

In this Section we assume that $S=(S_{t}^{(i)},i=1,...N;0\leq t\leq T)$ is a
multidimensional gBm and $R_{t}=rT$.

\textbf{5.1 Kirk's type approximation for basket-spread option. }A
basket-spread option price for $N$ assets is defined by the formula%
\begin{equation}
C_{T}=\mathbb{E}\left(
e^{-rT}\sum_{i=1}^{M}S_{T}^{(i)}-e^{-rT}\sum_{i=M+1}^{N}S_{T}^{(i)}-e^{-rT}K%
\right) ^{+},  \label{basketprice}
\end{equation}%
which is a particular case of (\ref{eqn:AsianBasketPrice}).

The first analytical formula for $C_{T}~$\ was obtained in \cite%
{margrabe1978value}) for the two-asset exchange option with $M=1$, $N=2$ and 
$K=0$ under gBm by making a reduction to the Black-Scholes formula
(Margrabe's formula). As with Asian options, it is not yet possible to
obtain an exact solution for this option under the Black-Scholes framework
when $N\geq 2~$and $K>0$, due to the fact that a linear combination of
log-normal random variables still has a non-identifiable distribution. These
options are traded in a variety of markets and are useful for hedging
portfolios of long and short positions in underlying assets (see \cite%
{carmona2003pricing}).

A number of different approaches to approximate the basket pricing problem
are found in the literature and they may be classed as numerical,
closed-form or semi-closed-form (see e.g. \cite{Krekel}). A popular
closed-form approximation was described in \cite{kirk1995correlation}
(Kirk's approximation).

Of particular interest is the recent paper by Lau and Lo \cite{LaoLo} which
extends Kirk's approximation to multi-asset baskets of $N>2$. We now briefly
describe their extension of Kirk's method.

Note that under the risk-neutral measure P the price of the option is given
by

\begin{equation*}
F_{t}=e^{-rt}\sum_{i=1}^{M}S_{t}^{(i)}
\end{equation*}
and 
\begin{equation*}
P_{t}=e^{-rt}\sum_{i=M+1}^{N}S_{t}^{(i)}+e^{-rT}K.
\end{equation*}%
Since $S_{t}^{(i)}$ are gBm the dynamics of $F_{t}$ and $P_{t}$ are given by 
\begin{align*}
dF_{t}& =\sum_{i=1}^{M}\sigma
_{i}S_{t}^{(i)}dB_{t}^{(i)}=F_{t}\sum_{i=1}^{M}\sigma _{i}\frac{S_{t}^{(i)}}{%
F_{t}}dB_{t}^{(i)}, & & F_{0}=\sum_{i=1}^{M}S_{0}^{(i)}, \\
dP_{t}& =\sum_{i=M+1}^{N}\sigma
_{i}S_{T}^{(i)}dB_{t}^{(i)}=P_{t}\sum_{i=M+1}^{N}\sigma _{i}\frac{S_{t}^{(i)}%
}{P_{t}}dB_{t}^{(i)} & & P_{0}=\sum_{i=M+1}^{N}S_{0}^{(i)}+e^{-rT}K,
\end{align*}%
where $\sigma _{i}$ are the volatilities of$~S_{t}^{(i)}$ and $B_{t}^{(i)}$
are correlated standard Brownian motions with correlation coefficients%
\begin{equation*}
\rho _{ij}=E(B_{1}^{(i)}B_{1}^{(j)}).
\end{equation*}%
Kirk's approximation, and its extension to the multi-dimensional case in 
\cite{LaoLo}, are based on an approximation of $%
(S_{t}^{(i)}/F_{t},S_{t}^{(i)}/P_{t})$ for $0\leq t\leq T~$by corresponding
constants $(S_{0}^{(i)}/F_{0},S_{0}^{(i)}/P_{0})$ , so $F_{t}$ is then
approximated by the process $\tilde{F}_{t}$ with dynamics%
\begin{equation*}
d\tilde{F}_{t}=\tilde{F}_{t}\sum_{i=1}^{M}\tilde{\sigma}_{i}dB_{t}^{(i)},\ 
\tilde{\sigma}_{i}=\sigma _{i}\frac{S_{0}^{(i)}}{F_{0}}.
\end{equation*}%
Similarly we may approximate $P_{t}~$by $\hat{P}_{t}$ with dynamics%
\begin{equation*}
d\hat{P}_{t}=\hat{P}_{t}\sum_{i=M+1}^{N}\hat{\sigma}_{i}dB_{t}^{(i)},\ \hat{%
\sigma}_{i}=\sigma _{i}\frac{S_{0}^{(i)}}{P_{0}+e^{-rT}K}.
\end{equation*}%
Using $\tilde{F}_{T}$ and $\hat{P}_{T}$ the option price (\ref{basketprice})
is then approximated by 
\begin{equation}
\mathbb{E}\left( \tilde{F}_{T}-\hat{P}_{T}\right) ^{+},  \label{eq:4}
\end{equation}%
which is precisely of the form for which Magrabe's formula applies since $%
\tilde{F}_{T}$ and $\hat{P}_{T}$ are lognormal. As reported in \cite{LaoLo},
this heuristic method gives accurate results for a range of $N$~and $M$,
however there are no analytical formulas for the accuracy of the
approximations made so far.

\textbf{5.2 Lower bounds. }We now apply the methods described above in
Section 2 to find a lower bound for the basket-spread option price. Using
the random variable $X_{T}:=(\ln \tilde{F}_{T}-\ln \hat{P}_{T})$ as our
proxy in (\ref{LB}), we have the following lower bound%
\begin{equation}
LB^{BC}:=\sup_{z\in \mathbb{R}}\left( 
\begin{array}{c}
\sum_{i=1}^{M}e^{-rT}\mathbb{E}\mathbb{I}\left\{ X_{T}>z\right\} S_{T}^{(i)}
\\ 
-\sum_{i=M+1}^{N}e^{-rT}\mathbb{E}\mathbb{I}\left\{ X_{T}>z\right\}
S_{T}^{(i)}-e^{-rT}K\mathbb{P}(X_{T}>z)%
\end{array}%
\right) .  \label{LB BC}
\end{equation}%
Next note that 
\begin{equation*}
X_{T}=\sum_{i=1}^{M}\tilde{\sigma}_{i}B_{T}^{(i)}-\sum_{i=M+1}^{N}\hat{\sigma%
}_{i}B_{T}^{(i)}+const
\end{equation*}%
is normally distributed, so $\mathbb{P}(X_{T}>z)$ is easy to compute for
each $z$. In order to evaluate the expectations $\mathbb{E}\mathbb{I}\left\{
X_{T}>z\right\} S_{T}^{(i)}~$in (\ref{LB BC}) observe that for every $i\in
\{1,\cdots ,N\}$ 
\begin{equation}
\mathbb{E}\mathbb{I}\left\{ X_{T}>z\right\} S_{T}^{(i)}=S_{0}^{(i)}\mathbb{%
\tilde{P}}^{(i)}(X_{T}>z)  \label{eq:6}
\end{equation}%
where the measures $\mathbb{\tilde{P}}^{(i)}$ are defined via the Girsanov
transformation with the Radon-Nikodym derivative 
\begin{equation*}
\frac{d\mathbb{\tilde{P}}^{(i)}}{d\mathbb{P}}=\exp \left\{ \sigma
_{i}B_{T}^{(i)}-T\sigma _{i}^{2}/2\right\} .
\end{equation*}%
It is known that the Girsanov transformation preserves covariances of
Gaussian random processes (see e.g. \cite{NKLfBm}, Proposition 2) and
therefore we have%
\begin{equation*}
\mathbb{\tilde{E}}^{(i)}\tilde{B}_{1}^{(i)}\tilde{B}_{1}^{(j)}=\rho _{ij}.
\end{equation*}%
Observe also that under the measure $\mathbb{\tilde{P}}^{(i)}$ 
\begin{equation*}
\left\{ \tilde{B}_{t}^{(j)}:=B_{t}^{(j)}-\rho _{ij}\sigma _{i}t,~0\leq t\leq
T,~j=1,2,\ldots ,M\right\}
\end{equation*}%
are standard Brownian motions. Thus we can easily find $\mathbb{\tilde{P}}%
^{(i)}(X_{T}>z)$ giving the lower bound 
\begin{equation*}
LB^{BC}=e^{-rT}\sup_{z\in \mathbb{R}}\left( 
\begin{array}{c}
\sum_{i=1}^{M}S_{0}^{(i)}\mathbb{\tilde{P}}^{(i)}(X_{T}>z) \\ 
-\sum_{i=M+1}^{N}S_{0}^{(i)}\mathbb{\tilde{P}}^{(i)}(X_{T}>z)-K\mathbb{P}%
(X_{T}>z)%
\end{array}%
\right) ,
\end{equation*}%
where $\mathbb{\tilde{P}}^{(i)}(X_{T}\leq z)$ and $\mathbb{P}(X_{T}\leq z)$
are Gaussian distributions with known parameters.

\textbf{5.3 Numerical results. }The above considerations can be easily
adapted to the case where each stock $S_{t}^{(i)}$ has a constant dividend
denoted $q_{i}$; the formulas above still hold except we replace $%
S_{0}^{(i)}e^{q_{i}T}$.

Tables 4 and 5 show results for 3 and 200 asset basket options with
parameter values from \cite{LaoLo} to allow for comparisons.

1) For the 3 asset case the parameters are $N=3$, $M=0$, $T=1$, $\sigma
_{1/2/3}=0.3$, $r=0.05$, $S_{0}=(60,50,40)$, $q_{1/2/3}=0.03$, $\rho
_{12}=0.7$, $\rho _{13}=0.5$ and $\rho _{23}=0.3$;

2) The parameter values for the 200 asset case are $N=100$, $M=50$, $T=1$, $%
\sigma _{i\leq M}=0.5$,$\sigma _{i>M}=0.3$, $r=0.05$, $S_{i\leq M}=10$, $%
S_{i>M}=9$, $q_{i\leq M}=0.05$, $q_{i>M}=0.03$, $\rho _{i\leq M\cap j\leq
M}=0.6$, $\rho _{i\leq M\cap j>M}=0.4$ and $\rho _{i\>M\cap j>M}=0.5$. 
\begin{table}[h]
\caption{ Results for 3 asset basket for varying strike $K$}
\label{tbl:Ling3Asset}\centering
\begin{tabular}{lcccc}
K & $LB^{BC}$ & Lau \& Lo & MC & SE \\ 
50.0 & 98.0054 & 98.0054 & 98.0235 & 0.0116 \\ 
100.0 & 50.9571 & 50.9654 & 50.968 & 0.0114 \\ 
150.0 & 15.7622 & 15.767 & 15.7787 & 0.008 \\ 
200.0 & 2.9862 & 2.9715 & 2.9999 & 0.0037 \\ 
250.0 & 0.4303 & 0.4222 & 0.4379 & 0.0014 \\ 
300.0 & 0.0551 & 0.0528 & 0.0565 & 0.0005 \\ 
\end{tabular}%
\end{table}

\bigskip

\begin{table}[h]
\caption{ Results for 200 asset basket for varying strike $K$}
\label{tbl:Ling200Asset}\centering
\begin{tabular}{lcccc}
K & $LB^{BC}$ & Lau \& Lo & MC & SE \\ 
0.0 & 143.937 & 143.937 & 144.023 & 0.0715 \\ 
50.0 & 119.833 & 119.834 & 119.9215 & 0.0671 \\ 
100.0 & 99.3342 & 99.3511 & 99.3813 & 0.0625 \\ 
150.0 & 82.0757 & 82.123 & 82.1487 & 0.058 \\ 
\end{tabular}%
\end{table}
\begin{equation*}
\end{equation*}

Both our lower bound and the Lau \& Lo method give good results for these
particular examples, though we note that our method is better for
out-of-the-money options. Both methods cannot give the exact price but with
the use of lower bounds we always know which side of the price that we are
on.

\bigskip

\textbf{6. Conclusions. }In this paper a framework for pricing average-type
options via lower and upper bounds has been proposed. Section 2 introduces
some general notation that is flexible enough to represent a wide variety of
such options with examples given for Asian, basket and VWAP options under
both continuous and discrete monitoring.

In Section 3 we looked at the VWAP problem under the so-called Levy-volume
model. It is also argued that lower bound approximations for the prices of
these VWAP options can be derived via lower bounds for ordinary Asian
options, thus reducing the dimensionality of the problem. Examples for the
case when the stock price process is a gBm and the volume process a gamma
process show that the method is accurate for different levels of volatility.

In Section 4 a method for calculating lower bounds for Asian options for
under both continuous and discrete monitoring is presented. The method is a
development of the two-step method of Thomson \cite{Thom} and relies heavily
on Fourier methods for the calculations.

An approximation for the delta of the option is also introduced. Numerical
experiments show that the lower bounds are very accurate when compared to MC
approximations. Finally in Section 5 a lower bound for the price of
basket-spread options is proposed and compared with an approximation based
on the mean-field approximations developed in \cite{LaoLo}. This lower
bound, under gBm, is derived via a series of measure-changes and reduces the
dimensionality of the pricing problem. The lower bound has the advantage,
when compared to the mean-field approximation, in that one knows on what
side of the price the approximation resides.

\textbf{Acknowledgment.} The authors thank Juri Hinz, Alex Buryak and other
participants of the working seminar at National Australian Bank for useful
discussions of the presented results.

\end{document}